\documentstyle[12pt]{aipproc}

\begin{document}

\input epsf

\title{An Update on Cosmological Anisotropy in Electromagnetic Propagation}

\author{John P. Ralston$^*$ and Borge Nodland$^{\dagger}$}
\address{$^*$Department of Physics and Astronomy\\
and Institute for Theoretical and Computational Science\\
University of Kansas, Lawrence, Kansas 66044\\
$^{\dagger}$Department of Physics and Astronomy\\
and Rochester Theory Center for Optical Science and Engineering\\
University of Rochester, Rochester, New York 14627}

\maketitle

\begin{abstract}
We review evidence for a new phenomenon in the propagation of radio
waves across the Universe, an anisotropic rotation of the plane of
polarization not accounted for by conventional physics.
\newline\newline [Contributed to the Proceedings of the 7th
International Conference on the Intersections of Particle and Nuclear
Physics (Big Sky, Montana, 1997), to be published by the American
Institute of Physics (Editor: T. W. Donnelly).]
\end{abstract}

Radio waves propagating on cosmological distances provide 
an exceedingly sensitive laboratory to explore new phenomena.  
Recently \cite{nod}, we reported an indication of anisotropy 
extracted from polarization measurements taken on distant radio 
galaxies. Here we review that work, subsequent reactions, and some 
new studies. 

Galaxies monitored in the 100 MHz--GHz range are found to emit linearly
polarized radio waves. The emission mechanism is believed to be
synchrotron radiation. The observed plane of polarization does not
usually align with the symmetry axis of the source (whose orientation
angle is denoted $\psi$) \cite{cla}. For decades, this has been studied
in terms of Faraday rotation in the intervening medium. Faraday
rotation can be taken out in a model--independent way, because the
Faraday angle of rotation $\theta_i(\lambda)$ for wavelength $\lambda$
goes like $\mbox{RM} \lambda^2$, where RM is a constant depending on
the integrated plasma parameters and magnetic field along the line of
sight. Consistent linear dependence on $\lambda^2$ is indeed observed.
However, the data fit requires more: for each source ($i$), the fits
are given by $\theta_i(\lambda) = \mbox{RM}_i \lambda^2 + \chi_i$. The
de--rotated polarization (whose orientation angle is $\chi$) does not
generally align with the galaxy major axis, nor with the axis at 90
degrees to the major axis; statistics on the offset of angles have
puzzled astronomers for 30 years \cite{cla}.

Analysis of the data is challenging. The data set is small, 160
sources. The angular distribution on the sky is highly non--uniform, as
is the distribution in the distances $r(z)$ to the sources. Finally,
for decades astronomers have used a particular arbitrary difference,
$\chi - \psi$, as a reference variable. There is an underlying
conceptual issue, that both the galaxy axis and the polarizations are
projective variables that return to themselves after a rotation of
$\pi$ (not $2 \pi$). Straightforward use of $\chi-\psi$, binned between
arbitrary intervals and defined $\bmod \;\pi$, will obscure correlations
that may depend on the direction of travel of a wave; this is a crucial
requirement if one is going to test for anisotropy. It also does not
allow an analysis to keep the important distinction between an obtuse
angle and the complementary acute one. To keep track of the difference
between obtuse and acute angles, and to allow a test for directional
dependency, we created \cite{nod} the angles $\beta^+$ and $\beta^-$,
defined by

\begin{equation}
\beta^+ = 
\left\{ \begin{array}{ll}
\chi-\psi & \mbox{if $\chi-\psi\geq 0$}\\
\chi-\psi+\pi & \mbox{if $\chi-\psi < 0$}
\end{array}
\right.
\;\;\;\;\;
\beta^- = 
\left\{ \begin{array}{ll}
\chi-\psi-\pi & \mbox{if $\chi-\psi\geq 0$}\\
\chi-\psi & \mbox{if $\chi-\psi < 0.$}
\end{array}
\right.
\label{eq1}
\end{equation}

A test for directional dependency is allowed by defining the rotation
angle $\beta$ as $\beta=\beta^+$ when $\cos\gamma > 0$ and
$\beta=\beta^-$ when $\cos\gamma < 0$, where $\gamma$ is the angle
between a spatial direction $\vec{s}$ and the propagating wavevector of
the wave. We employed Monte Carlo methods to search for correlations
\cite{nod}. We made thousands of fake galaxies with random $\chi$ and
$\psi$, while keeping their positions on the dome of the sky the same
as those of the real galaxies. As we varied $\vec{s}$ systematically
across the dome of the sky, we calculated (for each $\vec{s}$) the
probabilistic P-value of linear correlations in the observed data
relative to the random sets (Procedure 1). In the full data set, we
found a correlation described by $\beta = (r/\Lambda_s) \cos\gamma$.
This indicates anisotropy. Making a cut on $z>0.3$, which selects the
most distant half of the data set, we found a striking correlation,
with probabilistic P-values that the observed correlation would be
produced by random angular fluctuations to be less than $10^{-3}$; the
effect is $3.7 \sigma$.

A separate study (Procedure 2) eliminated possible bias in Procedure
1's determination of $\vec{s}$. Procedure 2 determined the
$\vec{s}$--direction that yielded the highest correlation for a
specific random data set; the resulting ``best'' correlations were then
accumulated for over a thousand different random sets. We then
calculated the probabilistic P-value of finding the correlation of the
observed data set relative to the best correlated random sets so
constructed. This gave a P-value less than 0.006, corresponding to
$2.7 \sigma$. The fits to the parameters $\Lambda_s$ and $\vec{s}$ are
$\Lambda_s= (1.1 \pm 0.08) 10^{25} \frac{h_0}{h} \mbox{m}$ and
$\vec{s}=({\rm Decl.}, {\rm R.A.})^*_s = (0^\circ \pm 20^\circ, 21
\mbox{hrs} \pm 2 \mbox{hrs})$, where $h_0=\frac{2}{3} 10^{-10}
\mbox{(years)}^{-1}$ and $h$ is the Hubble constant. We do not find a
significant correlation for $z<0.3$; in our full data set we also do
not find a significant correlation of $\beta = (\mbox{const}) r$.

Other searches have been conducted looking for systematic rotation
depending only on distance $r(z)$. For example, Carroll, Field and
Jackiw (CFJ) \cite{car} looked at the same data set as we did. However,
CFJ used $\chi - \psi$, which mixes up obtuse and acute angles, and
doesn't allow for straightforward correlation analysis to be done.
Moreover, one could easily question any systematic correlation with
distance as possibly due to the evolution of a population: there was a
``two--population'' hypothesis 30 years ago \cite{cla}, surmising that
nearby sources emitted waves of polarization parallel to the sources'
major axes, and that distant sources emitted waves of polarization
perpendicular to the sources' major axes. We looked for an anisotropic
correlation on the sky firstly because we had a theory that predicted
it \cite{nod}, and secondly because no population hypothesis could
reasonably explain a signal if seen.

We have concentrated mainly on data analysis, because establishing that
the effect is statistically significant is the first step. As for
theoretical questions, in Ref. \cite{nod} we noted a gauge invariant
modification of electrodynamics which has the same number of parameters
as needed to explain the data, and is compatible with precision
measurements of the Standard Model. Another possible explanation might
be a domain wall from an axion--like particle condensate. A third
explanation might be the twisting of polarization predicted by Brans
\cite{bran}, and Matzner and Tolman \cite{sup}, from parallel transport
in ordinary general relativity in an anisotropic cosmology. Several
alternate cosmologies or theories of gravity \cite{sup} have also come
to our attention, with claims that the anisotropic effect we saw would
(or, in some cases, was) predicted.

Not unexpectedly, after publication of Ref. \cite{nod} there has been
considerable controversy. Early charges that the data was ``old and
incomplete'' were found to be spurious, as we have been vigorously
reassured that the data set we used is the most complete one available
\cite{kro}. The statistics have come under attack, but we have seen
nothing valid to alter our conclusions. Eisenstein and Bunn (EB)
\cite{eis} quickly criticized the work on the basis of a single scatter
plot of $\beta$ versus $r \cos\gamma$ included in Ref. \cite{nod}. EB
observed that the data for $z>0.3$ was not uniformly distributed along
the $\beta$--axis. Without making any calculations, but ``estimating by
eye'' as EB put it, the observed data set for $z>0.3$ was claimed not
to be better correlated than data with $\beta$--values randomly
distributed about 90 degrees. EB conjectured that the anisotropic
correlation would go away if one compared the observed data to data
with random shufflings of the observed $\beta$--values, rather than
comparing (as we did, and in fact, only in Procedure 1) the observed
data to data with $\beta$--values obtained from uniform random
distributions of $\chi$ and $\psi$. (The distributions of the observed
$\chi$ and $\psi$ are, in fact, uniform in the full data set).

There are several problems with EB's criticism, explained in more
detail in our response to EB in Ref. \cite{eis}: (1) Procedure 2, the
more demanding, was ignored; (2) The scatterplot showed data after a
cut; this, combined with the fact that the data was linearly
correlated, predicts the distribution in $\beta$ to be as observed; (3)
Since shuffled data from such a limited, cut region is pre--correlated,
comparison of this data with the observed data may underestimate the
statistical significance of a real correlation in the observed data;
(4) Since eyeball estimates can be deceiving, we actually did the ``EB
calculation:'' we found that our correlation survived, both in
identifying the same anisotropy axis as found in Ref. \cite{nod}, and
in being statistically significant, with a P-value less than $2 \times
10^{-2}$. This was also independently confirmed numerically by P. Jain,
who also reproduced our original numerical results in Ref. \cite{nod}.
[We note that a more appropriate way to do shuffling is to shuffle
the observed pairs ($\psi$, $\chi$), not the $\beta$'s; this resulted
in an even stronger signal. Furthermore, we shuffled only the $\beta$'s
and ($\psi$, $\chi$) pairs from the $z \geq 0.3$ set, although it is
more reasonable to use the total set for this purpose].

Carroll and Field (CF) recently released a preprint \cite{car2}, based
on the ``two--population hypothesis.'' Restricting themselves to
$z>0.3$, CF confirmed some of our calculations, and also restated their
previous calculations (in Ref. \cite{car}) using the variable
$\chi-\psi$. With this variable, they found no indication of a
correlation nor anisotropy. As explained above, the variable is
unsuitable (as also pointed out in our response to Leahy in Ref.
\cite{lea}), and the procedure used by CF will generally miss an
anisotropic signal even in perfectly correlated data sets.
Unfortunately, CF have not made a clear distinction between $\beta$ and
$\chi-\psi$ in their presentation, leading some readers to think that
their calculation applies to our work, which used the
direction-of-travel-dependent $\beta$. The same problem of not using
$\beta$ occurs in Leahy, Ref. \cite{lea}.

Other reactions \cite{lea,war} have come from radio astronomers, who so
far have not addressed the anisotropic correlations on the sky, which
was the basic nature of our result in Ref.\cite{nod}. Instead, these
studies use high resolution VLBI data examining special internal
structures of selected objects, such as jets in the detailed radio maps
of certain nice--looking quasars. There are some problems with the
claims: (1) their data seems to be highly selected, with much smaller
sets than we used, and ignoring parts of the quasars which are not
``pretty;'' (2) some of the logic appears to be circular, depending
strongly on theoretical ideas engineered to understand the same data
before birefringence was considered; (3) the methodology lacks
statistical criteria. If these problems are ignored, the claims state
that no polarization rotation of the size we found is seen in the VLBI
data. One actually goes beyond the scope of Ref. \cite{nod} when one
compares the VLBI data and our data in this way: we reported an
anisotropic correlation in our data, along with a thorough statistical
characterization of its likelihood, and also pointed out the
possibility that it was caused by systematic bias in our data.

Naturally, we believe that the VLBI data contains great potential for
information, and should be studied systematically on its own basis. By
this, we mean that the VLBI data should be subjected to the same kind
of statistical Monte Carlo tests for anisotropy as the tests we applied
to our data. One cannot deduce much from any particular polarization
rotation found in one or a few galaxies. Much more relevant would be
statistical significance of a possible anisotropic signal, as quantifed
by the P--values obtained from analyses like ours.

It is also interesting to note that the frequency of the VLBI data is
much higher than the frequency of the data we studied, so perhaps the
studies in Refs. \cite{lea,war} really are looking at different
phenomena. For example, we would think that the Brans mechanism
\cite{bran} (if it exists) is independent of frequency. Regarding other
hypotheses, there will be other predictions. For example, what is known
generally about the frequency dependence of domain walls? It would seem
to be model dependent to compare the different types of data in Refs.
\cite{lea,war} and Ref. \cite{nod} prematurely.

Several recent articles report interesting progress. Obhukov et al.
\cite{obu} claim that the correlation we observed could be caused by
global rotation. K\"{u}hne \cite{kuh}, and Bracewell and Eshleman
\cite{bra} have independently observed that the anisotropy axis
extracted in Ref. \cite{nod} coincides tolerably well with the
direction of the cosmic microwave background (CMB) dipole axis. This
seems to be a strong clue, although our calculations on kinematic
Doppler effects on Faraday rotation do not yield any mechanism to
suggest a connection. It thus seems possible that some non--kinematic,
grand medium effect may be at work, but we don't know what.

\begin{figure}
\centerline{
\epsfxsize= \textwidth
\epsfbox[0 530 500 715]{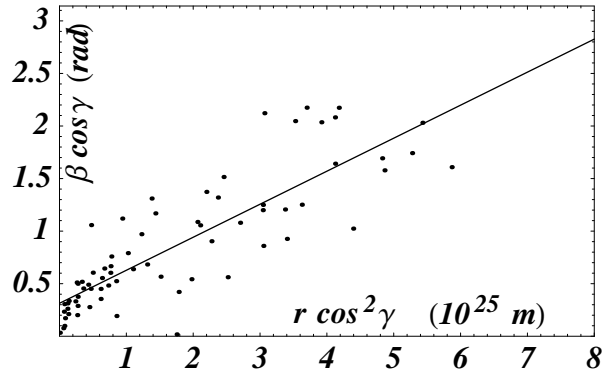} }
\caption{
Plot of $\beta\cos\gamma$ versus $r \cos^2 \gamma$, for the data of
Fig. 1(d) in Ref. \protect\cite{nod}. This plot is equivalent to that
of Ref. \protect\cite{nod}, but places all data points in the first
quadrant. A definite correlation can be seen.}
\label{fig1}
\end{figure}

Finally, some critics seemed to be uncomfortable with the fact that our
data for $\beta$ versus $r \cos \gamma$ occupied both the first and
third quadrants. We have taken this to heart by plotting
$\beta\cos\gamma$ versus $r \cos^2 \gamma$ instead, which brings all
the data into the first quadrant. The Jacobian of this transformation
is a power of $\cos\gamma$, scaling both of the variables of interest
by the same function. The plot is shown in Fig. 1, for the data of Fig.
1(d) in Ref. \cite{nod}. It must be noted that ``eyeballing'' scatter
plots is dangerous; the plot in these variables happen to look nice,
but one should rely on quantitative statistical measures. Yet the
figure may make it more clear to the eye that there is a definite
signal of anisotropy. The next question, yet unanswered, is: why?


\begin{references}

\bibitem{nod} B. Nodland and J. P. Ralston, {\it Phys. Rev.
Lett.} {\bf 78}, 3043 (1997); preprint astro-ph/9704196.

\bibitem{cla} J. N. Clarke {\it et al.}, {\it Mon. Not. R. Astron. Soc.}
{\bf 190}, 205 (1980); F. F. Gardner and J. B. Whiteoak, {\it Nature}
{\bf 197}, 1162 (1963).

\bibitem{car} S. M. Carroll, G. B. Field and R. Jackiw, {\it Phys.
Rev.} D {\bf 41}, 1231 (1990).

\bibitem{bran} C. H. Brans, {\it Astrophys. J.} {\bf 197}, 1 (1975).

\bibitem{sup} R. A. Matzner and B. W. Tolman, {\it Phys. Rev. D} {\bf
26}, 2951 (1982); M. Sachs, {\it General Relativity and Matter}
(Reidel, 1982); R. B. Mann and J. W. Moffat, {\it Can. J. Phys.} {\bf
59} 1730 (1981); C. M. Will, {\it Phys. Rev. Lett.} {\bf 62}, 369
(1989); R. B. Mann, J. W.  Moffat, and J. H. Palmer, {\it ibid} {\bf
62}, 2765 (1989); J. W. Moffat, preprint astro-ph/9704300 (1997); D.
V. Ahluwalia and T. Goldman, {\it Mod. Phys. Lett.} {\bf A28}, 2623
(1993); Y. N. Obukhov, V. A. Korotky and F. W. Hehl, preprint
astro-ph/9705243 (1997); A. Dobado and A. L. Maroto, preprint
astro-ph/9706044 (1997).

\bibitem{kro} P. Kronberg, private communication.

\bibitem{eis} D. J. Eisenstein and E. F. Bunn, preprint
astro-ph/9704247 (1997); B. Nodland and J. P. Ralston, preprint
astro-ph/9705190 (1997).

\bibitem{car2} S. M. Carroll and G. B. Field, preprint astro-ph/9704263
(1997).

\bibitem{lea} J. P. Leahy, preprint astro-ph/9704285 (1997); B. Nodland
and J. P. Ralston, preprint astro-ph/9706126 (1997).

\bibitem{war} J. F. C. Wardle, R. A. Perley, and M. H. Cohen, 
preprint astro-ph/9705142 (1997).

\bibitem{obu} Y. N. Obukhov, V. A. Korotky and F. W. Hehl, preprint
astro-ph/9705243 (1997).

\bibitem{kuh} R. W. K\"{u}hne, private communication and preprint
submitted to {\it Phys. Rev. Lett.} (1997).

\bibitem{bra} R. N. Bracewell and V. R. Eshleman,
preprint aps1997jun13\_006 (1997).

\end{references}
\end{document}